\documentclass[aps,prb,twocolumn,showpacs,superscriptaddress]{revtex4}
\usepackage{amssymb,amsmath}
\usepackage{graphicx}
\usepackage{dcolumn}
\usepackage{bm}

\begin{document}

\title{The Limit of Mechanical Stability in Quantum Crystals: \\
       A Diffusion Monte Carlo Study of Solid $^{4}$He}

\author{Claudio Cazorla}
\affiliation{Institut de Ci$\grave{e}$ncia de Materials de 
             Barcelona (ICMAB-CSIC), 08193 Bellaterra, Spain}
\author{Jordi Boronat}
\affiliation{Departament de F\'{i}sica i Enginyeria Nuclear, 
             Universitat Polit\`{e}cnica de Catalunya, Campus 
             Nord B4-B5, E-08034, Barcelona, Spain}
\email{ccazorla@icmab.es}

\begin{abstract}
We present a first-principles study of the energy and elastic properties
of solid helium at pressures below the range in which is energetically 
stable. We find that the limit of mechanical stability in hcp $^{4}$He 
is $P_{s} = -33.82$~bar, which lies significantly below the spinodal 
pressure found in the liquid phase (i.e., $-9.6$~bar). Furthermore, we 
show that the pressure variation of the transverse and longitudinal  
sound velocities close to $P_{s}$ do not follow a power law of the 
form $\propto \left( P - P_{s} \right)^{\gamma}$, in contrast to what 
is observed on the fluid.     
\end{abstract}
\pacs{67.80.-s,02.70.Ss,67.40.-w}
\maketitle

\section{Introduction}
\label{sec:introduction}

In the last two decades, extensive theoretical and experimental works have 
focused on the study of liquid helium at negative pressures and ultra-low 
temperatures.~\cite{nissen89,maris93} The equation of state of this 
material has been measured very accurately within the density range in which 
is stable, and extrapolated to the region of negative pressures.~\cite{maris95,maris02} 
On the theoretical side, quantum Monte Carlo methods have allowed for precise 
and explicit simulation of metastable liquid helium, producing results which 
are in remarkable good agreement with experiments.~\cite{boronat94a} A quantity 
of central interest in all these studies is the spinodal pressure $P_{s}$, that is, 
the pressure at which the bulk modulus, $B$, and sound velocity vanish and 
thus the liquid becomes unstable against long wavelength density fluctuations. 
In liquid $^{4}$He $P_{s}$ amounts to $-9.6$~bar, while in liquid $^{3}$He to 
$-3.2$~bar.~\cite{maris95,maris02}   

Analogous studies performed in solid helium are almost non-existent in the 
literature.~\cite{maris09,maris10} At $T = 0$~K solid helium becomes 
stable at pressures larger than $\sim 25$~bar, thus certainly the realm of 
negative pressures lies well below such a threshold. Nevertheless, analysis of the 
mechanical instability limit in helium crystals turns out to be a topic of 
fundamental interest. In contrast to liquids, solids can sustain shear stresses 
and because of this ordinary fact the definition of the spinodal density in 
crystals differs greatly from the one given above. In the particular
case of solid $^{4}$He, previous attempts to determine $P_{s}$ have been, 
to the best of our knowledge, only tentative.~\cite{maris09,maris10}  

The energy of a crystal under a homogeneous elastic deformation is 
\begin{equation}
E(V,\eta) = E_{0}(V) + \frac{1}{2} V \sum_{ij} C_{ij} \eta_{i} \eta_{j}~,
\label{eq:energy}
\end{equation}
where $\lbrace C_{ij} \rbrace$ are the elastic constants and $\lbrace \eta_{i} \rbrace$
a general strain deformation (both expressed in Voigt notation). The specific 
symmetry of a crystal determines the number of independent elastic constants 
which are different from zero. In the case of hexagonal crystals (e.g., hcp $^{4}$He) 
these are five: $C_{11}$, $C_{12}$, $C_{13}$, $C_{33}$ and $C_{44}$. The 
conditions for mechanical stability in a crystal follow from the requirement 
that upon a general strain deformation the change in the total energy must
be positive. It can be shown that in hcp crystals subjected to an external 
pressure $P$, these conditions readily are~\cite{sinko02,grimvall12}
\begin{eqnarray}
C_{44} - P &>& 0~[{\rm C1}]~ \nonumber \\
C_{11} - C_{12} - 2P &>& 0~[{\rm C2}]~ \nonumber \\
\left( C_{33} - P \right) \left( C_{11} + C_{12} \right) - 2 \left( C_{13} + P \right)^{2} &>& 0~[{\rm C3}]~. 
\label{eq:conditions}
\end{eqnarray}
The spinodal pressure, or limit of mechanical stability, in a crystal then is 
identified with the point at which any of the three conditions above is not fulfilled. 
We must note that the bulk modulus of an hexagonal crystal can be expressed as~\cite{cazorla12}
\begin{equation}
B = -V \frac{dP}{dV} = \frac{C_{33} \left( C_{11} + C_{12}\right)-2C_{13}^{2}}{C_{11} + C_{12} + 2C_{33} - 4C_{13}}~,
\label{bulkmodulus}
\end{equation}
thus the conditions at which the bulk modulus (and sound velocities) vanishes in 
general must not coincide with the corresponding limit of mechanical stability. 
This reasoning is in fact very different from the usual $P_{s}$ analysis performed 
in liquids.   

In this article we present a computational study of the energetic and elastic
properties in solid $^{4}$He at pressures below $\sim 25$~bar, based on the 
diffusion Monte Carlo method. Our first-principles calculations allow us to 
determine with precision the limit of mechanical stability in this 
crystal, which we estimate to be $P_{s} = -33.82$~bar (that is, much larger
in absolute value than the one found in the fluid). Moreover, we show 
that, in contrast to liquid helium, the pressure variation of the sound velocities 
near $P _{s}$ do not follow a power law of the form $\propto \left( P - P_{s} \right)^{\gamma}$ 
with $\gamma = 1/3$.\cite{maris91} 

The organization of this article is as follows. In the next section, we provide the 
details of our computational method and calculations. In Sec.~\ref{sec:results}, we 
present our results and comment on them. Finally, we summarize our main findings in 
Sec.~\ref{sec:conclusions}.

\section{Computational Details}
\label{sec:methods}
In DMC the time-dependent Schr\"odinger equation 
of a $N$-particle system is solved stochastically by 
simulating the time evolution of the Green's function 
propagator $e^{-\frac{i}{\hbar} \hat{H} t}$ in imaginary time 
$\tau \equiv \frac{it}{\hbar}$. In the $\tau \to \infty$ limit, sets of 
configurations (walkers) $\lbrace {\bf R}_i \equiv {\bf r}_1,\ldots,{\bf r}_N \rbrace$ 
render the probability distribution function $\Psi_0 \Psi$, 
where $\Psi_0$ is the true ground-state wave function and $\Psi$ 
a guiding wave function used for importance sampling. Within DMC, virtually 
exact results (i.e., subjected to statistical uncertainties only) can be 
obtained for the ground-state energy and related quantities.~\cite{barnett91,casulleras95} 

We are interested in studying the ground-state of hcp $^4$He, which 
we assume to be governed by the Hamiltonian  
$H= -\frac{\hbar^2}{2m_{\rm He}} \sum_{i=1}^{N} \nabla_i^2 + \sum_{i<j}^{N} V_{\rm He-He}(r_{ij})$
where $m_{\rm He}$ is the mass of a $^{4}$He atom and $V_{\rm He-He}$ 
the semi-empirical pairwise potential due to Aziz.~\cite{aziz2} 
It is worth noticing that the Aziz potential provides an 
excellent description of the He-He interactions at low 
pressure.~\cite{boro94,cazorla08a} 

The guiding wave function that we use in this study, $\Psi_{\rm SNJ}$, reproduces 
both the crystal ordering and Bose-Einstein symmetry. This model wave function was  
introduced in Ref.~[\onlinecite{cazorla09b}] and reads 
\begin{equation}
\Psi_{\rm SNJ}({\bf r}_1,\ldots,{\bf r}_N) = \prod_{i<j}^{N} f(r_{ij}) 
\prod_{J=1}^{N} \left( \sum_{i=1}^{N} g(r_{iJ}) \right)~,
\label{snjtrial}
\end{equation}
where the index in the second product runs over the lattice
position vectors.
In previous works we have demonstrated that $\Psi_{\rm SNJ}$ 
provides an excellent description of the ground-state properties of 
bulk hcp $^{4}$He~\cite{cazorla09b} and quantum solid 
films.~\cite{cazorla08b,cazorla10,gordillo11}
Here, we adopt the correlation functions in Eq.~(\ref{snjtrial}) of the
McMillan, $f(r) = \exp\left[-1/2~(b/r)^{5}\right]$~, and Gaussian, 
$g(r) = \exp\left[-1/2~(a r^{2})\right]$, forms, and optimize 
the corresponding parameters with variational Monte Carlo (VMC) [in order
to simulate the metastable solid we performed one unique VMC optimization
at $P \sim 25$~bar].

The technical parameters in our calculations were set in order to 
ensure convergence of the total energy per particle to less than 
$0.02$~K/atom. The value of the mean population 
of walkers was $500$ and the length of the imaginary 
time-step ($\Delta \tau$) $5 \cdot 10^{-4}$~K$^{-1}$~. 
We used large simulation boxes containing $200$ atoms in all the 
cases. Statistics were accumulated over $10^{5}$ DMC steps
performed after system equilibration, and the approximation used 
for the short-time Green's function $e^{-\hat{H} \tau}$ is accurate 
to second order in $\tau$.~\cite{boronat94,chin90} 
 
The computational strategy that we followed for calculation of the 
elastic constants $\lbrace C_{ij} \rbrace$ is the same than explained 
in Refs.~[\onlinecite{cazorla12,cazorla13,cazorla12b}], thus we address 
the interested reader to them.

\section{Results and Discussion}
\label{sec:results}

\begin{figure}
\centerline
        {\includegraphics[width=1.0\linewidth]{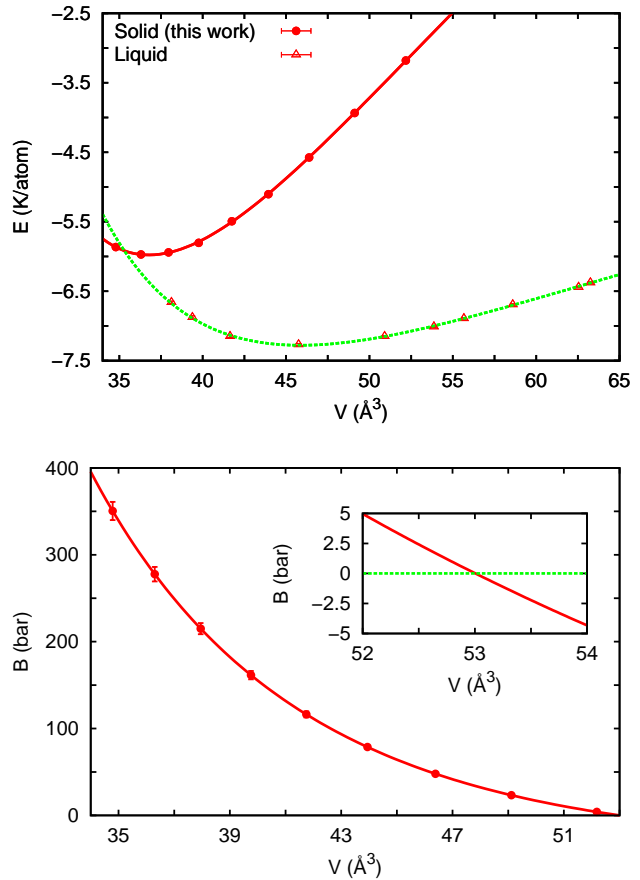}}
\caption{({\emph Top:})~Calculated energy per particle expressed as
         a function of volume. The solid line represents a polynomial
         fit to our results (see text).
         Results obtained in the liquid phase (from work [\onlinecite{boronat94a}])
         are shown for comparison. ({\emph Bottom:})~Calculated bulk modulus
         expressed as a function of volume. The region in which $B$ vanishes
         is augmentated on the inset.}
\label{eos}
\end{figure}

In Fig.~\ref{eos}, we show the calculated energy in solid helium 
expressed as a function of volume (solid symbols). We fitted 
our results to the polynomial curve  
\begin{equation}
E (V) = E_{0} + b \left[ \left( \frac{V_{0}}{V} \right) - 1 \right]^{2} + c \left[ \left( \frac{V_{0}}{V} \right) - 1 \right]^{3}~,
\label{eq:dmcenergy}
\end{equation} 
and found as best parameters $E_{0} = -5.978(5)$~K, $V_{0} = 36.90(5)$~\AA$^{3}$, 
$b = 34.38(5)$~K, and $c = 7.86(5)$~K. In comparison to the analogous energy curve 
obtained in the liquid (see Fig.~\ref{eos}),~\cite{boronat94a} $E (V)$ displays 
larger slopes, which translates into larger negative pressures, at volumes close to 
equilibrium (i.e.,  $V_{0}$). In the same figure, we enclose the bulk modulus, 
$B (V) = V d^{2}E/dV^{2}$, that is calculated directly from $E (V)$ (solid line). 
The symbols which appear therein correspond to estimations obtained with Eq.(\ref{bulkmodulus}), 
thereby excellent consistency between our energy and $C_{ij}$ calculations is demonstrated. 
We find that the bulk modulus of helium vanishes at volume $V_{B} = 53.00$~\AA$^{3}$ and 
pressure $P_{B} = -34.06$~bar. In what follows, we present our analysis of the elastic 
properties and mechanical stability in solid $^{4}$He and clarify whether $V_{B}$ and $P_{B}$ 
are meaningful physical quantities.     

\begin{figure}
\vspace{2.00cm}
\centerline
        {\includegraphics[width=1.0\linewidth]{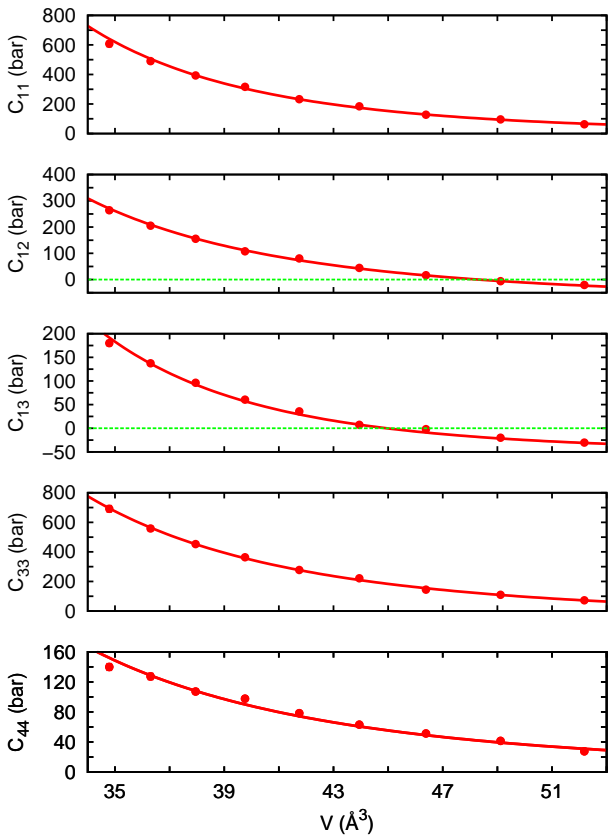}}
\vspace{-2.75cm}
\caption{Calculated elastic constants expressed as a function of 
         volume. The solid lines are power-law fits to our results
         (solid symbols).}
\label{cij}
\end{figure}

\begin{figure}
\centerline
        {\includegraphics[width=1.0\linewidth]{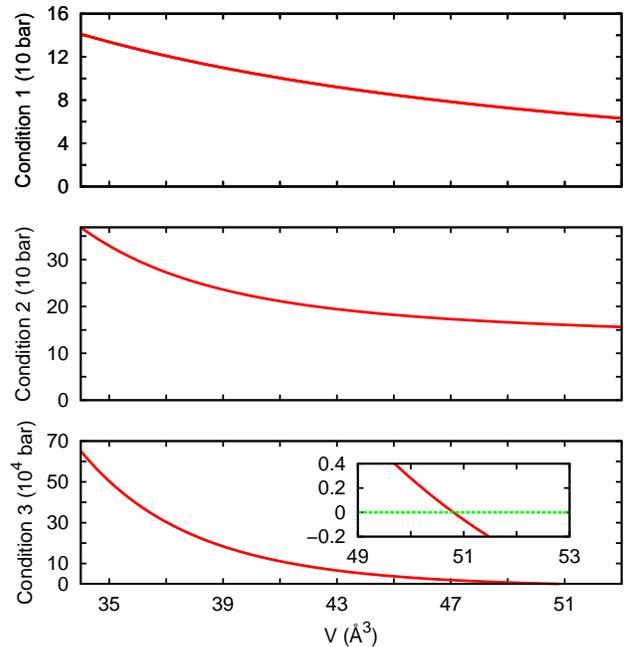}}
\caption{Conditions of mechanical stability [see Eq.(\ref{eq:conditions})] 
         expressed as a function of volume. The region in which condition~3
         is not accomplished is augmentated on the inset.}
\label{mechcond}
\end{figure}

Fig.~\ref{cij} shows how the calculated elastic constants in solid helium 
change as a function of volume. As it is
appreciated therein, all five $C_{ij}$ decrease with increasing volume.
In particular, $C_{12}$ and $C_{13}$ become zero at $V = 48.32$~($P = -32.80$~bar)
and $44.95$~\AA$^{3}$~($-29.47$~bar). We performed power-law fits to our $C_{ij} (V)$ 
results in order to render continuous elastic constant functions. By doing this, 
we were able to represent the three conditions of mechanical stability in hcp helium 
[see Eq.(\ref{eq:conditions})] within the volume interval of interest. These results 
are enclosed in Fig.~\ref{mechcond} and show that condition~$3$, i.e., 
[C3] in Eq.(\ref{eq:conditions}), is violated at $V_{s} = 50.81$~\AA$^{3}$ and 
$P_{s} = -33.82$~bar. Thus, we identify $( V_{s} , P_{s})$ with the limiting 
state at which solid $^{4}$He is mechanically stable. We note that these 
conditions, although similar, are not equal to $(V_{B} , P_{B})$ at which the bulk 
modulus vanishes [as it was already expected from Eqs.(\ref{eq:conditions})-(\ref{bulkmodulus})]. 

\begin{figure}
\centerline
        {\includegraphics[width=1.0\linewidth]{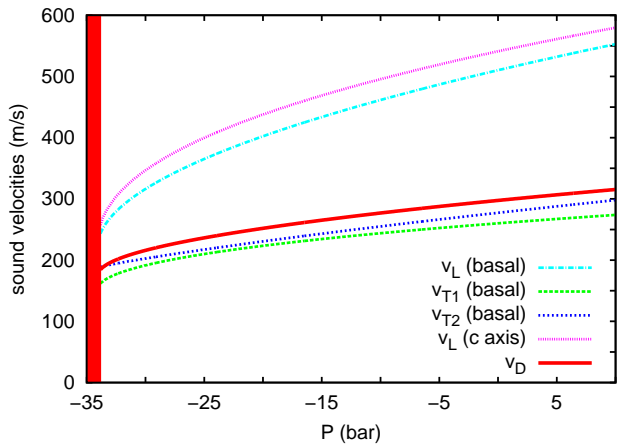}}
\caption{Calculated sound velocities along the $c$-axis and in the basal plane
         [see Eqs.(\ref{eqn:cvel})-(\ref{eqn:basalvel})] expressed as 
         a function of pressure. The region in which the crystal becomes 
         mechanically unstable is filled with red color.}
\label{soundvel}
\end{figure}

Sound velocities in solids can be either longitudinal or tranverse and depend 
on the direction of propagation. In crystals with hexagonal symmetry two 
main propagation modes are identified, one along the $c$-axis and the 
other contained in the basal plane. The relationships between the 
elastic constants and sound velocities in hcp crystals are~\cite{cazorla12}
\begin{eqnarray}
v_{L} &=& \left( C_{33} / \rho \right)^{1/2}~\nonumber \\
v_{T1} = v_{T2}  &=& \left( C_{44} / \rho \right)^{1/2}
\label{eqn:cvel}
\end{eqnarray}
along the $c$-axis, and
\begin{eqnarray}
v_{L} &=& \left( C_{11} / \rho \right)^{1/2}~\nonumber   \\ 
v_{T1} = \left( C_{11} - C_{12} / 2 \rho \right)^{1/2} &\quad& v_{T2} = \left( C_{44} / \rho \right)^{1/2}
\label{eqn:basalvel}
\end{eqnarray}
in the basal plane. In Fig.~\ref{soundvel}, we plot the six sound velocities 
calculated in solid helium as a function of pressure (some of them are coincident) 
together with an appropriate average of them, $v_{D}$.~\cite{cazorla12} We 
observe that none of the sound velocities vanishes at the spinodal pressure $P_{s}$ 
(marked in red in the figure), as it was already expected from 
Eqs.(\ref{eq:conditions})-(\ref{eqn:cvel})-(\ref{eqn:basalvel}).
Neither any sound velocity becomes zero at $P_{B}$, the pressure at
which the bulk modulus vanishes. These results differ greatly from 
the usual conclusions drawn on the mechanical stability in liquids. 

Recently, it has been suggested that the variation of the sound velocities near the 
spinodal density in solid $^{4}$He could follow a power law of the form 
$\propto \left( P - P_{s}\right)^{\gamma}$ where $\gamma = 1/3$,
in analogy to what is observed in the liquid.~\cite{maris09,maris10} 
However, in light of the results shown in Fig.~\ref{soundvel} such an hypotheses  
must be rejected since none of the sound velocities becomes zero at $P_{s}$. 
Furthermore, we performed fits of the form $v_{L,T} (P) = a + b \left( P - P_{s}\right)^{\beta}$
to our results and found that parameter $\beta$ was always different from $1/3$.  
For instance, in the $v_{D}$ case we found $\beta = 0.55$ as the optimal value.  
Therefore, we must conclude that the propagation of sound waves in helium crystals at the 
verge of mechanical instability differs radically from that predicted in liquid helium 
under similar conditions. 

\section{Conclusions}
\label{sec:conclusions}
We have presented a rigorous and precise computational study of the energy and elastic
properties of solid helium at conditions in which is metastable. We have determined 
the limit of mechanical stability in this crystal, i.e., $P_{s} = -33.82$~bar, and 
the variation of the sound velocities at pressures close to it. Overall, we demonstrate
that solid and liquid helium behave radically different in the vicinity of their 
spinodal state limits.

\begin{acknowledgments}
This work was supported by MICINN-Spain [Grants No. MAT2010-18113,
CSD2007-00041, FIS2011-25275, and CSIC JAE-DOC program (C.C.)], 
and Generalitat de Catalunya [Grant No.~2009SGR-1003]. 
\end{acknowledgments}


\begin{thebibliography}{30}
\bibitem{nissen89} J. A. Nissen, E. Bodegom, L. C. Brodie, J. S. Semura, 
                   Phys. Rev. B \textbf{40}, 6617 (1989).
\bibitem{maris93} H. J. Maris, S. Balibar, M. S. Pettersen, 
                  J. Low Temp. Phys. \textbf{93}, 1069 (1993).
\bibitem{maris95} H. J. Maris, J. Low Temp. Phys. \textbf{98}, 403 (1995).
\bibitem{maris02} H. J. Maris, D. O. Edwards, J. Low Temp. Phys. \textbf{129}, 1 (2002).
\bibitem{boronat94a} J. Boronat, J. Casulleras, and J. Navarro, 
                     Phys. Rev. B \textbf{50}, 3427 (1994).
\bibitem{maris09} H. J. Maris, J. Low Temp. Phys. \textbf{155}, 290 (2009).
\bibitem{maris10} H. J. Maris, J. Low Temp. Phys. \textbf{158}, 485 (2010).
\bibitem{sinko02} G. V. Sin'ko and N. A. Smirnov, 
                  J. Phys.: Condens. Matter \textbf{14}, 6989 (2002).
\bibitem{grimvall12} G. Grimvall \emph{et al.}, Rev. Mod. Phys. \textbf{84}, 945 (2012).
\bibitem{cazorla12} C. Cazorla, Y. Lutsyshyn, and J. Boronat, 
                    Phys. Rev. B \textbf{85}, 024101 (2012).
\bibitem{maris91} H. J. Maris, Phys. Rev. Lett. \textbf{66}, 45 (1991).
\bibitem{barnett91} R. Barnett, P. Reynolds, and W. A. Lester Jr., J. Comput. Phys.
                    \textbf{96}, 258 (1991).
\bibitem{casulleras95} J. Casulleras and J. Boronat, Phys. Rev. B \textbf{52}, 3654 (1995).
\bibitem{aziz2} R. A. Aziz, F. R. W. McCourt, and C. C. K. Wong, Mol. Phys.
                \textbf{61}, 1487 (1987).
\bibitem{boro94} J. Boronat and J. Casulleras, Phys. Rev. B \textbf{49}, 8920 (1994).
\bibitem{cazorla08a} C. Cazorla and J. Boronat, 
                     J. Phys.: Condens. Matter \textbf{20}, 015223 (2008).
\bibitem{cazorla09b} C. Cazorla, G. Astrakharchick, J. Casulleras, and J. Boronat,
                     New Journal of Phys. \textbf{11}, 013047 (2009).
\bibitem{cazorla08b} C. Cazorla and J. Boronat, Phys. Rev. B \textbf{77}, 024310 (2008).
\bibitem{cazorla10}  C. Cazorla, G. Astrakharchick, J. Casulleras, and J. Boronat,
                     J. Phys.: Condens. Matter \textbf{22}, 165402 (2010).
\bibitem{gordillo11} M. C. Gordillo, C. Cazorla, and J. Boronat,
                     Phys. Rev. B \textbf{83}, 121406(R) (2011).
\bibitem{boronat94} J. Boronat and J. Casulleras, Phys. Rev. B \textbf{49}, 8920 (1994).
\bibitem{chin90} S. A. Chin, Phys. Rev. A \textbf{42}, 6991 (1990).
\bibitem{cazorla13} C. Cazorla, Y. Lutsyshyn, and J. Boronat,
                    Phys. Rev. B \textbf{87}, 214522 (2013).
\bibitem{cazorla12b} R. Rota, Y. Lutsyshyn, C. Cazorla, and J. Boronat, 
                     J. Low Temp. Phys. \textbf{168}, 150 (2012). 
\end{thebibliography}
\end{document}